\documentclass[envcountsame]{llncs}
\usepackage{pscproc2}
\usepackage{graphics}

\usepackage{xspace,amsmath,url}
\usepackage{color}
\usepackage{booktabs}

\usepackage{hyperref}

\usepackage{graphics}
\usepackage{epstopdf}
\usepackage{epsfig}

\newcommand{\ceil}[1]{{\lceil #1 \rceil}}
\newcommand{\floor}[1]{{\lfloor #1 \rfloor}}

\definecolor{red}{rgb}{1.0,0,0}

\def\optp{k}
\def\setP{\ensuremath{{\mathcal K}}}

\def\jitdesc{MAG (short of Multi AOSO on $q$-Grams)\xspace}
\def\nonjitdesc{SMAG (short of Simple Multi AOSO on $q$-Grams)\xspace}

\def\jit{MAG\xspace}
\def\nonjit{SMAG\xspace}

\newcommand{\uand}{\mathrel{\&}}

\begin{document}

\title{Multiple pattern matching revisited}

\author{Robert Susik$^\dag$, Szymon Grabowski$^\dag$, Kimmo Fredriksson$^\ddag$}
\institute{$^\dag$ Lodz University of Technology, 
           Institute of Applied Computer Science,\\
           Al.\ Politechniki 11, 90--924 {\L}\'od\'z, Poland, 
           \email{\{rsusik|sgrabow\}@kis.p.lodz.pl}\\
           $^\ddag$ School of Computing, University of Eastern Finland, \\
           P.O.B. 1627, FI-70211 Kuopio, Finland, \email{kimmo.fredriksson@uef.fi}
}

\maketitle

\begin{abstract}
We consider the classical exact multiple string matching problem. Our solution is
based on $q$-grams combined with pattern superimposition, bit-parallelism and
alphabet size reduction. We discuss the pros and cons of the various alternatives
of how to achieve best combination. Our method is closely related to previous
work by (Salmela et al., 2006). The experimental results show that our method
performs well on different alphabet sizes and that they scale to large pattern
sets.
%
\end{abstract}


\section{Introduction}

Multiple pattern matching is a classic problem, with about 40 years of history, 
with applications in intrusion detection, anti-virus software and bioinformatics, 
to name a few.
The problem can be stated as follows: 
Given text $T$ of length $n$ and pattern set 
$\mathcal{P} = \{P_1, \ldots, P_r\}$, in which each pattern 
is of length $m$, and all considered sequences are over 
common alphabet $\Sigma$ of size $\sigma$, find all pattern occurrences in $T$.
The pattern equal length requirement may be removed.
The multiple pattern matching problem is a straightforward 
generalization of single pattern matching 
and it is no surprise that many techniques worked out for a single pattern 
are borrowed in efficient algorithms for multiple patterns.

\subsection{Related work}

The classical algorithms for the present problem can be roughly divided into three different 
categories, ($i$) prefix searching, ($ii$) suffix searching and ($iii$) factor searching. 
Another way to classify the solutions is to say that they are based on character comparisons,
hashing, or bit-parallelism. Yet another view is to say that they are based on filtering, aiming
for good average case complexity, or on 
some kind of ``direct search'' with good worst case complexity guarantees.
These different categorizations are of course not mutually exclusive, and many solutions are
hybrids that borrow ideas from several techniques.
For a good overview of the classical solutions we refer the reader e.g.\ to 
\cite{NRbook2002,Gus1997,CHL2007}.
We briefly review some of them in the following.

Perhaps the most famous solution to the multiple pattern matching 
problem is the Aho--Corasick (AC)~\cite{AC1975} algorithm, 
which works in linear time (prefix-based approach).
It builds a pattern trie with extra (failure) links and actually 
generalizes the Knuth--Morris--Pratt algorithm \cite{KMP1977} for a single pattern.
More precisely, AC total time is $O(M + n + z) = O(M + n)$, where $M$, the 
sum of pattern lengths, is the preprocessing cost, and $z$ is 
the total number of pattern occurrences in $T$.
Recently 
Fredriksson and Grabowski~\cite{FG2009} showed an average-optimal 
filtering variant of the classic AC algorithm.
They built the AC automaton over superimposed subpatterns, 
which allows to sample the text characters in regular distances, 
not to miss any match (i.e., any verification).
This algorithm is based on the same ideas as the current work.

Another classic algorithm is Commentz--Walter~\cite{CW1979}, which generalizes
the ideas of Boyer--Moore (BM) algorithm \cite{BM77} for a single pattern to solve 
the multiple pattern matching problem (suffix-based approach). 
Set Horspool~\cite{FV2001,NRbook2002} may be considered its more practical 
simplification, exactly in the way that Boyer--Moore--Horspool (BMH) \cite{Horspool80}
is a simplification of the original BM.
Set Horspool makes use of a generalized bad character function.
The Horspool technique was used in a different way in an earlier 
algorithm by Wu and Manber~\cite{WM1994}. These methods are based on backward matching over
a sliding text window, which is shifted based on some rule, and the hope is that many
text characters can be skipped altogether.

The first factor based algorithms were DAWG-match \cite{DBLP:journals/ipl/CrochemoreCGLPR99} 
and MultiBDM \cite{CR1994}. Like 
Commentz--Walter and Set Horspool they are based on backward matching. However, 
instead of recognizing the pattern suffixes, they recognize the factors, which effectively
means that they work more per window, but in return they are able to make longer shifts of 
the sliding window, and in fact they obtain optimal average case complexity.
At the same time they are linear in the worst case. The drawback is that these algorithms
are reasonably complex and not very efficient in practice. More practical approach is
the Set Backward Oracle Matching (SBOM) algorithm \cite{AR99}, 
which is based on the same idea as MultiBDM,
but uses simpler data structures and is very efficient in practice. 
Yet another variant is the Succinct Backward DAWG Matching algorithm (SBDM) \cite{Fjea09}, which 
is practical for huge pattern sets due to replacing the suffix automaton with succinct index.
The factor based algorithms usually lead to average optimal \cite{NF2004} 
complexity $O(n \log_\sigma (rm) / m)$.

Bit-parallelism can be used to replace the various automata in the previous methods to obtain
very simple and very efficient variants of many classical algorithms. The classic method for 
a single pattern is Shift-Or \cite{BYG92}. The idea is to encode (non-deterministic) automaton
as a bitvector, i.e.\ a small integer value, and simulate all the states in parallel 
using Boolean logic and arithmetic.
The result is often the most practical method for the problem, but the drawback is that the
scalability is limited by the number of bits in a computer word, although there exist ways to 
alleviate this problem somewhat, see \cite{PT03,DBLP:journals/iandc/CantoneFG12}.
Another way that is applicable to huge pattern sets is to combine bit-parallelism with
$q$-grams; our method is also based on this, and we review the idea and related previous work
in detail in the next section.

Some recent work also recognizes the neglected power of the SIMD instructions, which have been 
available on commodity computers well over a decade.
For example, Faro and K{\"u}lekci~\cite{FK2012} make use of the Intel 
Streaming SIMD Extensions (SSE) technology, which gives wide registers and many special purpose
instructions to work with. They develop (among other things) a 
\textsf{wsfp} ({\it word-size fingerprint instruction}) operation, based on hardware opcode
for computing CRC32 checksums, 
which computes an $\alpha$-bit fingerprint from a $w$-bit register 
handled as a block of $\alpha$ characters.
Similar values are obtained for all $\alpha$-sized factors of all the patterns 
in the preprocessing, and \textsf{wsfp} can therefore be used as a simple yet efficient
hash-function to identify text blocks that may contain a matching pattern.

The paper is organized as follows.
Section~2 describes and discusses the two key concepts underlying our work,
$q$-grams and pattern superimposition.
Section~3 presents the description of our algorithm, together with its
complexity analysis.
Section~4 contains (preliminary) experimental results.
The last section concludes and points some avenues for pursuing further
research.


\section{On $q$-grams and superimposition}

A $q$-gram is (usually) a contiguous substring (factor) of $q$ characters of a string, 
although non-contiguous $q$-grams have been considered \cite{DBLP:journals/fuin/BurkhardtK03}. 
In what follows,
$q$ can be considered a small constant, 2 \ldots 6 in practice, although we may analyze
the optimal value for a given problem instance.
We note that $q$-grams have been widely used in approximate (single and multiple) 
string matching, where they can be used to obtain fast filtering algorithms based on 
exact matching of a set of $q$-grams.
Obviously these algorithms work for the exact case as well, as a special case, but they are
not interesting in our point of view. Another use (which is not relevant in our case) 
is to speed up exact matching of a single pattern by treating
the $q$-grams as a superalphabet, see \cite{F03c}.

In our case $q$-grams are interesting as combined with a technique called superimposition.
Consider a set of patterns $\mathcal P = \{P_1, \ldots, P_r\}$. We form a single pattern $P$
where each position $P[i]$ is no longer a single character, but a {\em set} of characters,
i.e.\ $P[i] \subseteq \Sigma$. More precisely, $P[i] = \bigcup_j P_j[i]$. Now $P$ can be used
as a {\em filter}: we search candidate text substrings that might contain an occurrence of 
any of the patterns in $\mathcal P$. That is, if $T[i+j] \in P[j]$, for all $j \in 1 \ldots m$,
then $T[i \ldots i + m - 1]$ may match with some pattern in $\mathcal P$.

For example, if $\mathcal P = \{abba, bbac\}$, the superimposed pattern will be 
$P = \{a,b\} \{b\} \{a, b\} \{a, c\}$,
and there are a total of 8 different
strings of length 4 that can match with $P$ (and trigger verification).
Therefore we immediately notice
one of the problems with this approach, i.e.\ the probability that some text character $t$
matches a pattern character $p$ is no longer $1/\sigma$ (assuming uniform random distribution), 
it can be up to $r/\sigma$. This gets quickly out of hands when the number of patterns $r$ grows.

To make the technique more useful, we first generate a new set of patterns, and then superimpose.
The new patterns have the $q$-grams as the alphabet, which mean the new alphabet has size
$\sigma^q$, and the probability of a false positive candidate will be considerably lower.
There a two main approaches: overlapping and non-overlapping $q$-grams. 

Consider first the 
overlapping $q$-grams. For each $P_i$ we generate a new pattern such that 
$P'_i[j] = P_i[j \ldots j + q - 1]$, for $j \in 1 \ldots m - q + 1$, that is, each
$q$-gram $P_i[j \ldots j + q - 1]$ is treated as a single ``super character'' in $P'_i$. 
Note also that the pattern lengths are decreased from $m$ to $m - q + 1$.
Taking the previous example, if $\mathcal P = \{abba, bbac\}$ and now $q = 2$, the new pattern
set is $\mathcal P' = \{[ab][bb][ba], [bb][ba][ac]\}$, where we use the brackets to denote 
the $q$-grams. The corresponding superimposed pattern is then 
$P' = \{[ab],[bb]\}\{[bb],[ba]\}\{[ba],[ac]\}$.
To be able to search for $P'$, the text must be factored in exactly the same way.

The other possibility is to use non-overlapping $q$-grams. In this case we have
$P'_i[j] = P_i[(j-1)q + 1 \ldots jq]$, for $j \in 1 \ldots \floor{m/q}$, and for our 
running example we get 
$P' = \{[ab],[bb]\}\{[ba],[ac]\}$.
Again, the text must be factored 
similarly. But the problem now is that only every $q$th text position is considered,
and to solve this problem we must consider all $q$ possible shifts of the original 
patterns. That is, given a pattern $P_i$, we generate a set 
$\hat P_i = \{P_i[1 \ldots m], P_i[2 \ldots m,] \ldots, P_i[q-1 \ldots m\}$, and then generate
$\hat P'_i$, and finally superimpose them. 

The above two alternatives both have some benefits and drawbacks. For overlapping
$q$-grams we have:
\begin{itemize}
 \item pattern length is large ($m-q+1$), which means less verifications
 \item text length is practically unaffected  ($n-q+1$)
\end{itemize}
Non-overlapping:
\begin{itemize}
 \item pattern length is short ($m/q$), which means potentially more verifications, but bit-parallelism works for bigger $m$
 \item text is shorter too ($n/q$)
 \item more patterns to superimpose (factor of $q$)
\end{itemize}
In the end, the benefits and drawbacks between the two approaches mostly cancel out each other,
except for bit-parallelism remains more applicable to non-overlapping $q$-grams.


To illustrate the power of this technique, let us have, for example, a random text over an 
alphabet of size $\sigma = 16$ and patterns generated according to the same probability 
distribution; $q$-grams are not used yet (i.e., we assume $q$ = 1).
If $r = 16$, then the expected size of a character class in the superimposed 
pattern is about $10.3$, which means that a match probability 
for a single character position is about $64\%$.
Even if high, this value may yet be feasible for long enough patterns, 
but if we increase $r$ to 64, the character class expected size grows to 
over 15.7 and the corresponding probability to over $98\%$.
This implies that match verifications are likely to be invoked for most 
positions of the text.
Using $q$-grams has the effect of artificially growing the alphabet.
In our example, if we use $q = 2$ and thus $\sigma' = 16^2 = 256$, 
the corresponding probabilities for $r = 16$ and $r = 64$ become 
about $6\%$ and $22\%$, respectively, so they are significantly lower.

The main problem that remains is to decide between the two choices, properly choose a suitable
$q$, and finally find a good algorithm to search the superimposed pattern. 
To this end, Salmela et al.~\cite{STK2006} presented three algorithms 
combining the known mechanisms: Shift-Or, BNDM \cite{NRjea00} and BMH, with overlapping $q$-grams; 
the former two of these algorithms are bit-parallel ones. The resulting algorithm
were called SOG, BG and HG, respectively. 
In general larger $q$ means better filtering, but on the other hand the size of the data 
structures (tables) that the algorithms use is $O(\sigma^q)$, which can be prohibitive. BGqus 
\cite{yangetal} tries to solve the problem by combining BG with hashing.

In general, not many classic algorithms can be
generalized to handle superimposed patterns (character classes) efficiently, but bit-parallel
methods generalize trivially. In the next section we describe our choice, FAOSO \cite{FG2009}.

\section{Our algorithm}

In \cite{FG2009} a general technique of how to skip text
characters, with any (linear time) string matching algorithm that can search for multiple 
patterns simultaneously was presented, alongside with several applications to know algorithms.
In the following we review the idea, and for the moment assume that we already have done
all factoring to $q$-grams, and that we have only a single pattern.

\subsection{Average-optimal character skipping}

The method takes a parameter $\optp$, and from the original pattern generates a set
$\setP$ of $\optp$ new patterns $\setP=\{P^0, \ldots,
P^{\optp-1}\}$, each of length $m'=\floor{m/\optp}$, as follows:
\[
P^j [i] = P[j+i\optp], ~~~ j = 0 \ldots \optp-1, ~~~ i = 0 \ldots \floor{m/\optp}-1.
\]
In other words, $\optp$ different alignments of the original
pattern $P$ is generated, each alignment containing only every $\optp$th character. The
total length of the patterns $P^j$ is $\optp\floor{m/\optp} \leq m$.  

Assume now that $P$ occurs at $T[i \ldots i+m-1]$. From the
definition of $P^j$ it directly follows that
\[
P^j[h] = T[i+j+h\optp], ~~~ j = i\mod \optp, ~~~ h = 0 \ldots m'-1.
\]
This means that the set $\setP$ can be used as a filter for the pattern
$P$, and that the filter needs only to scan every $\optp$th character of
$T$.

The occurrences of the patterns in $\setP$ can be searched for simultaneously 
using any multiple string matching algorithm. 
Assuming that the selected string matching algorithm runs generally in $O(n)$ 
time, then the filtering time becomes $O(n/\optp)$, as only every $\optp$th symbol
of $T$ is read. 
The filter searches for the exact matches of $\optp$ patterns, each of length 
$\floor{m/\optp}$. Assuming that each character occurs with probability
$1/\sigma$, the probability that $P^j$ occurs (triggering a verification) 
in a given text position
is $(1/\sigma)^\floor{m/\optp}$. A brute force verification cost is in the
worst case $O(m)$.
To keep the total time
at most $O(n/\optp)$ on average, we select $\optp$ so that 
$nm/\sigma^{m/\optp} = O(n/\optp)$. This is satisfied for 
$\optp = m/(2\log_\sigma(m))$, where the verification cost 
becomes $O(n/m)$ and filtering cost $O(n \log_\sigma(m)/m)$.
The total average time is then dominated by the filtering time, i.e.\ 
$O(n \log_\sigma(m) / m)$, which is optimal \cite{Yao79}.

\subsection{Multiple matching with $q$-grams}

To apply the previous idea to multiple matching, we just assume that the (single) 
input pattern (for the filter) is 
the non-overlapping $q$-gram factored and superimposed pattern set. The verification phase 
just needs
to be aware that there are possibly more than one pattern to verify. The analysis remains 
essentially the same: now the text length is $n/q$, pattern lengths are $m/q$, there are $r$ 
patterns to verify, and the probability of a match is $p$ instead of $1/\sigma$, where
$p = O(1-(1-(1/\sigma^q))^{qr}) = O((qr)/\sigma^q)$. 
That is, the filtering time is $O(q n/(\optp q)) = O(n/\optp)$, verification 
cost is $O(rqm)$, and its probability is $O(p^{\floor{m/(\optp q)}})$ for each of the $n/q$
text positions. However, now we have two parameters to optimize, $\optp$ and $q$, and the optimal
value of one depends on the other.


In practice we want to choose $q$ first, such that the verification probability is as 
low as possible.
This means maximizing $q$, but the preprocessing cost (and space) grows as $O(\sigma^q)$, and 
we do not want this to exceed $O(rm)$ (or the filtering cost for that matter). So we select
$q = \log_\sigma (rm)$, and then choose $\optp$ as large as possible.
Repeating the above analysis gives then
\[
 \optp = O\left(\frac{m}{\log_\sigma(rm)} \cdot \frac{\log_\sigma 1/\rho}{\log_\sigma(rm)+\log_\sigma 1/\rho}\right),
\]
where $\rho = \log_\sigma(rm)/m$. We note that this is not average-optimal anymore, although
we are still able to skip text characters.



To actually search the superimposed pattern, we use FAOSO \cite{FG2009}, which is based on 
Shift-Or. The fact that the pattern consists of character classes is not a problem for 
bit-parallel algorithms, since it only affects the initial preprocessing of a single table.
For details see \cite{FG2009}. 
The filter implemented with FAOSO runs in $O(n/\optp \cdot \ceil{(m/q)/w})$ time in our case,
where $w$ is the number of bits in computer word (typically 64).

We note that Salmela et al.~\cite{STK2006} have tried a similar
approach, but dismissed it early because it did not look promising
for short patterns in their tests.

\subsubsection{Implementation.}

In the algorithms' point of view the $q$-gram, i.e.\ the super character, must have some
suitable representation, and the convenient way is to compute a numerical value in the 
range $0 \ldots \sigma^q - 1$, which is done as $\sum_{i=1}^q S[i] \cdot \sigma^{i-1}$ for 
a $q$-gram $S[1 \ldots q]$. This is computed using Horner's method to avoid the exponentiation.
We have experimented with two different variants. The first encodes the whole text prior to 
starting the actual search algorithm, which is then more streamlined. This also means
that the total complexity is $\Omega(n)$, the time to encode the text. We call the resulting
algorithm \nonjitdesc. The other alternative is to keep the text intact, and compute the numerical
representation of the $q$-gram requested on the fly. This adds just constant overhead to the 
total complexity. We call this variant \jitdesc. We have verified experimentally that 
\jit is generally better than \nonjit.

\subsection{Alphabet mapping}

If the alphabet is large, then selecting a suitable $q$ may become a problem. The reason is that
some value $q'$ may be too small to facilitate good filtering capability, yet, using $q=q'+1$ 
can be problematic, as the preprocessing time and space grow with $\sigma^q$ (note that $q$ must 
be an integer). The other view of using length $q$ strings as super characters, we may say that
our characters have $q \log_2 \sigma$ bits, and we want to have more control of how many bits
we use. One way to achieve this is to reduce the original alphabet size $\sigma$. 

We note that in theory this method cannot achieve much, as reducing the alphabet size generally
only worsens the filtering capability and therefore forces larger $q$, but in practice this 
allows better fine tuning of the parameters.

What we do is that we select some $\sigma' < \sigma$, compute a mapping 
$\mu: \Sigma \mapsto 0 \ldots \sigma' - 1$, and use $\mu(c)$ whenever the (filtering) 
algorithm needs to access some character $c$ from the text or the pattern set. Verifications
still obviously use the original alphabet. 

A simple method to achieve this is to compute the histogram of character distribution of the 
pattern set, and assign code $0$ to the most frequent character, $1$ to second most frequent,
and so on, and put the $\sigma'-1 \ldots \sigma -1$ most frequent characters to the last bin,
i.e.\ giving them code $\sigma'-1$. The text characters not appearing in the patterns also 
will have code $\sigma'-1$.

A better strategy is to try to distribute the original characters into $\sigma'$ bins so that
each bin will have (approximately) equal weight, i.e.\ each 
$\mu(c)$, where $c \in 0 \ldots \sigma'-1$
will have (approximately) equal probability of appearance. This is NP-hard optimization problem, so
we use a simple greedy heuristic.

\subsubsection{Alphabet mapping on the $q$-grams.}

We note that the above method can be applied also on the $q$-gram alphabet. This allows a precise
control of the table size, and combined with hashing, it can accommodate very large $q$ as well.
That is, we want to
\begin{enumerate}
\item	Choose some (possibly very large) $q$;
\item	compute the $q$-gram frequencies on the pattern set (using e.g.\ hashing to avoid 
possibly large tables);
\item	choose some suitable $\sigma'$, the size of the mapped $q$-gram alphabet;
\item	use method of choice (e.g.\ bin-packing) to reduce the number of
$q$-grams, i.e.\ map the $q$-grams to range $0...\sigma'-1$; 
\item	use hashing to store the mapping, along with the corresponding bitvectors needed by FAOSO.
\end{enumerate}

\subsubsection{Combined alphabet mapping and $q$-gram generation.}

Yet another method to reduce the alphabet is to combine the $q$-gram computations with some
bit magic. The benefit is that the mapping tables need not to be preprocessed, and this allows 
further optimizations as we will see shortly. The drawback is that the quality of the mapping 
is worse than what is achieved with approaches like bin-packing.

Consider a (text sub-)string $S[1 \ldots q]$ over alphabet $\Sigma$ of size $\sigma$. 
A simple way to reduce the alphabet is to consider only the $\ell$ low-order bits of each $S[i]$,
where $\ell < \log_2 \sigma$. We can then compute $q\ell$ bit $q$-gram $s$ simply as
\[
 s = (S[1] \uand b) + (S[2] \uand b) << \ell + (S[3] \uand b) << 2\ell + \ldots + (S[q] \uand b) << (q-1)\ell,
\]
where $b = (1 << \ell) - 1$ and $<<$ denotes the left shift and $\uand$ the bitwise and.

The main benefit of this approach is that a sequence of shifts and adds can be often replaced
by a multiplication (which can be seen as an algorithm performing just that). As an illustrative 
example, consider the case $\ell = 2$ and hence $b = 3$ (which coincides to DNA nicely). 
As an implementation detail, assume that the text is 8-bit ASCII text, and it is possible to 
address the text, a sequence of characters, 
as a sequence of 32 bit integers (which is easy e.g.\ in C). 
Then to compute a 8-bit 4-gram $s$ we can simply do
\[
 s = (((x >> 1) \uand 0x03030303) * 0x40100401) >> 24,
\]
where $x$ is the 32 bit integer containing the 4 chars $S[1 \ldots 4]$. Assuming 4 letter DNA 
alphabet, the right shift (by $1$) 
and the (parallel) masking generate 2-bit unique (and case insensitive) codes for all 4 characters.
If the alphabet is larger (some DNA sequences have rare additional symbols), those will be mapped
in the same range, $0 \ldots 3$. The multiplication then shifts and adds all those codes into a 8 bit 
quantity, and the final shift moves the 4-gram to the low order bits. 
Larger $q$-grams can be obtained by repeating the code.

We leave the implementation to future work.





\section{Preliminary experimental results}

In order to evaluate the performance of our approach, 
we run a few experiments, using the 200\,MB versions of 
selected datasets (\texttt{dna}, \texttt{english} and \texttt{proteins}) 
from the widely used Pizza~\&~Chili corpus 
(\url{http://pizzachili.dcc.uchile.cl/}).

We test the following algorithms:
\begin{itemize}
\item BNDM on $q$-grams (BG)~\cite{STK2006},
\item Shift-Or on $q$-grams (SOG)~\cite{STK2006},
\item BMH on $q$-grams (HG)~\cite{STK2006},
\item Rabin-Karp algorithm combined with binary search and two-level 
hashing (RK)~\cite{STK2006},
\item Multibom and Multibsom are variants of the 
Set Backward Oracle Matching algorithm \cite{AR99},
\item Succinct Backward DAWG Matching algorithm (SBDM) \cite{Fjea09},
\item Multi AOSO on $q$-Grams (MAG) (this work).
\end{itemize}

All codes were obtained from the original authors.
Our MAG was implemented in C++ and compiled with \texttt{g++} version 4.8.1
with \texttt{-O3} optimization.
The experiments were run on a desktop PC with an Intel i3-2100 CPU clocked 
at 3.1\,GHz with 128\,KB L1, 512\,KB L2 and 3 MB\,L3 cache.
The test machine was equiped with 4\,GB of 1333\,MHz DDR3 RAM 
and running Ubuntu 64-bit OS with kernel 3.11.0-17.

\begin{figure}[pt]
\centerline{
\includegraphics[width=0.49\textwidth,scale=1.0]{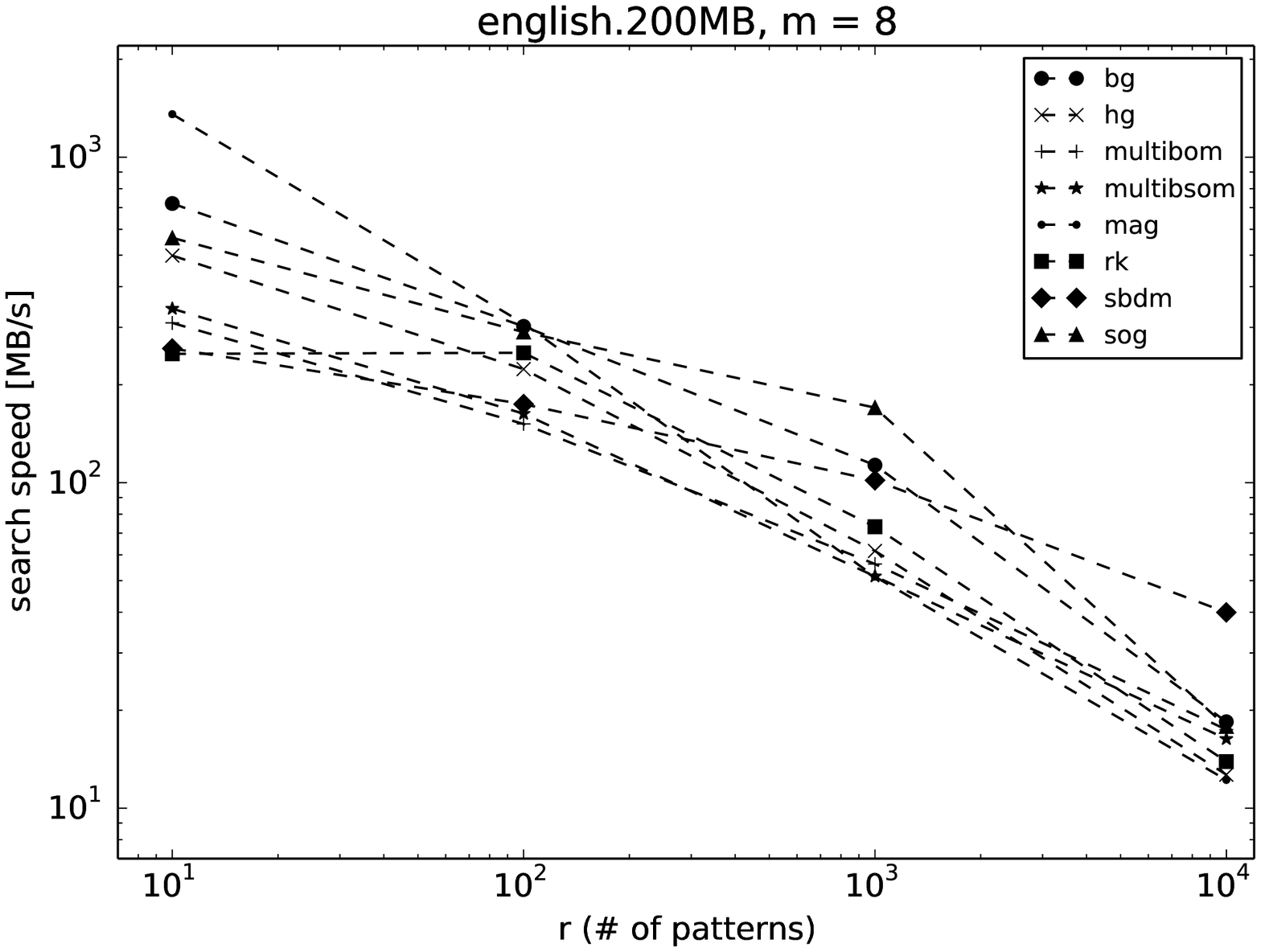}
\includegraphics[width=0.49\textwidth,scale=1.0]{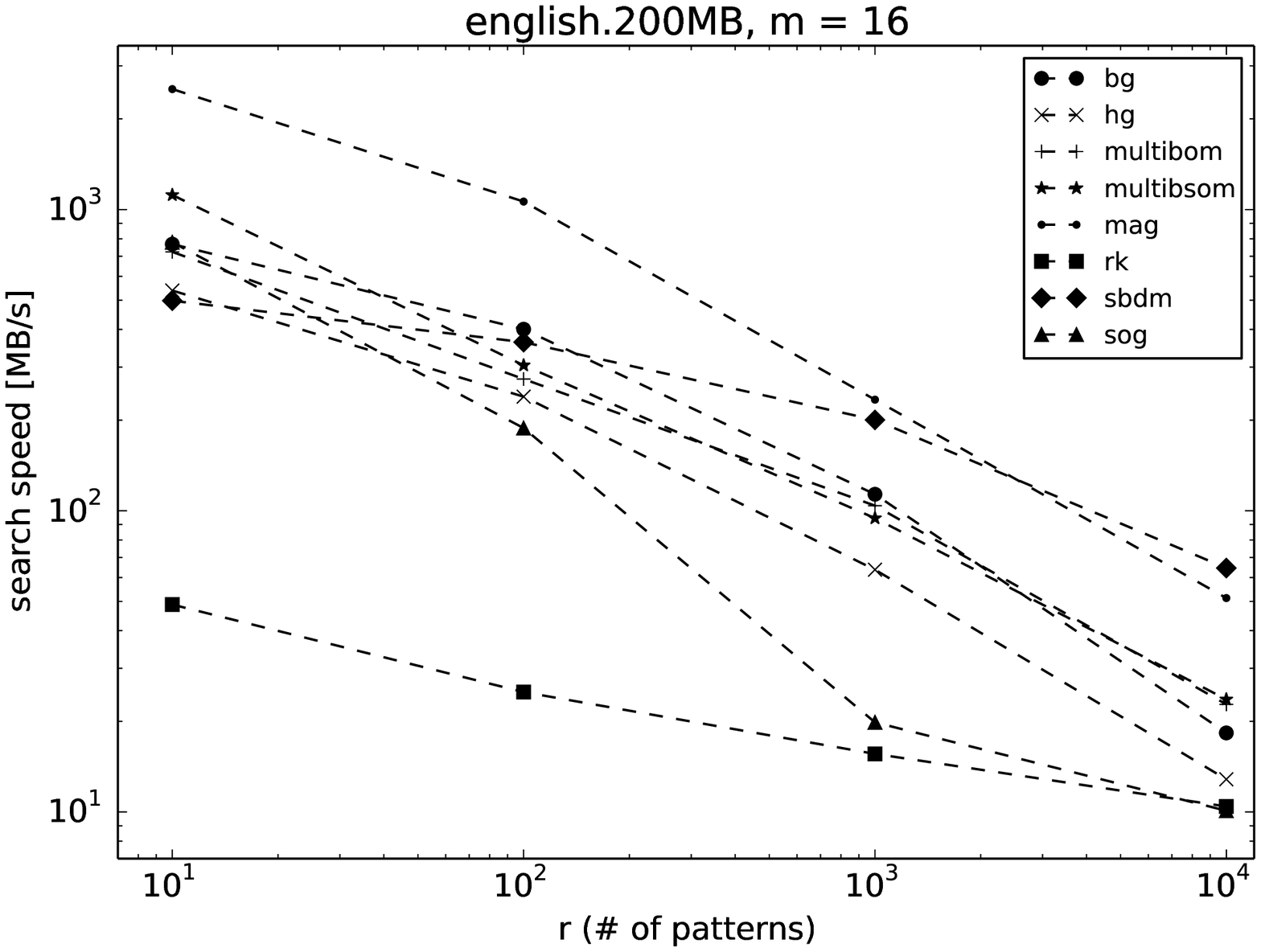}
}
\centerline{
\includegraphics[width=0.49\textwidth,scale=1.0]{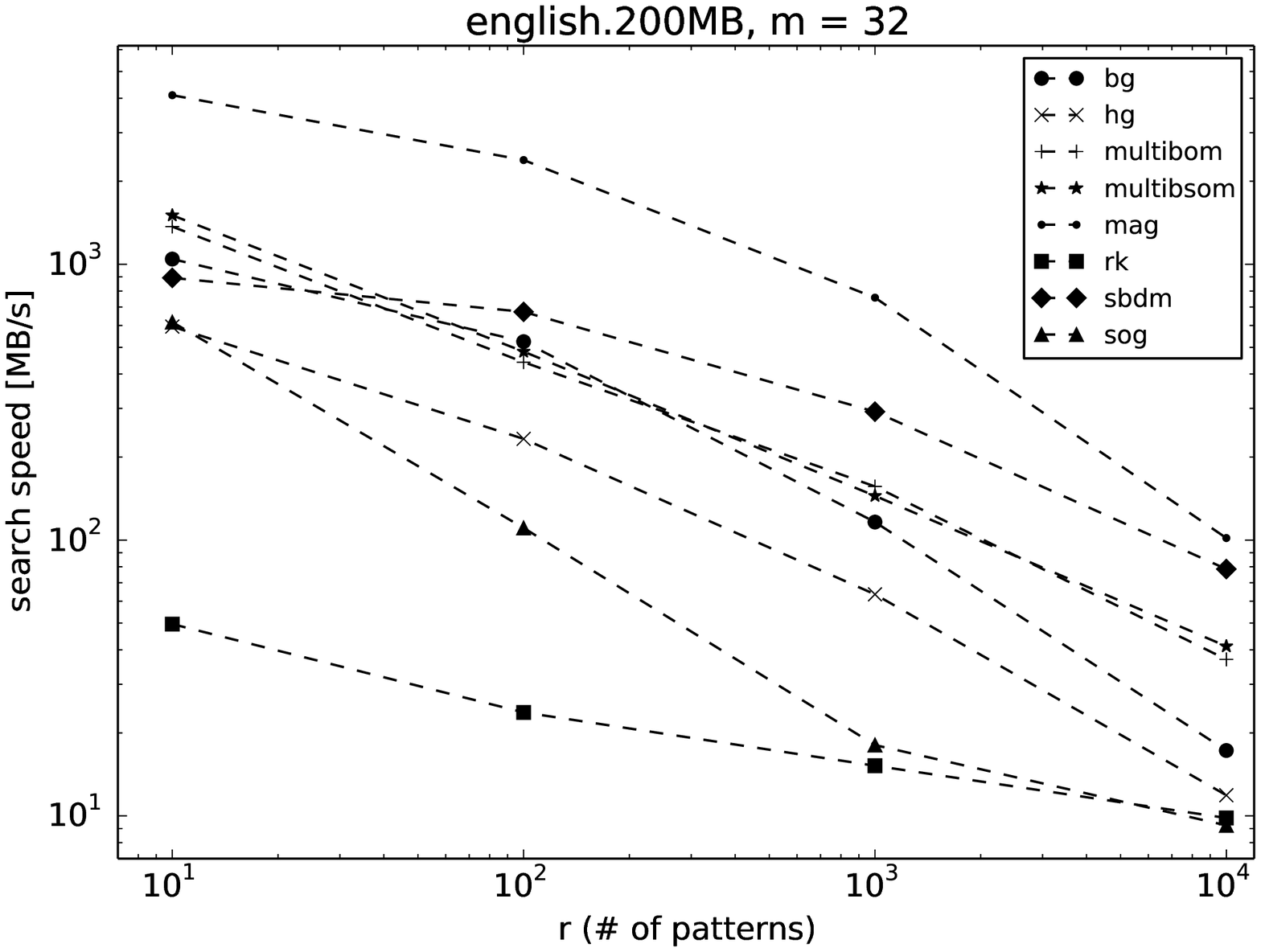}
\includegraphics[width=0.49\textwidth,scale=1.0]{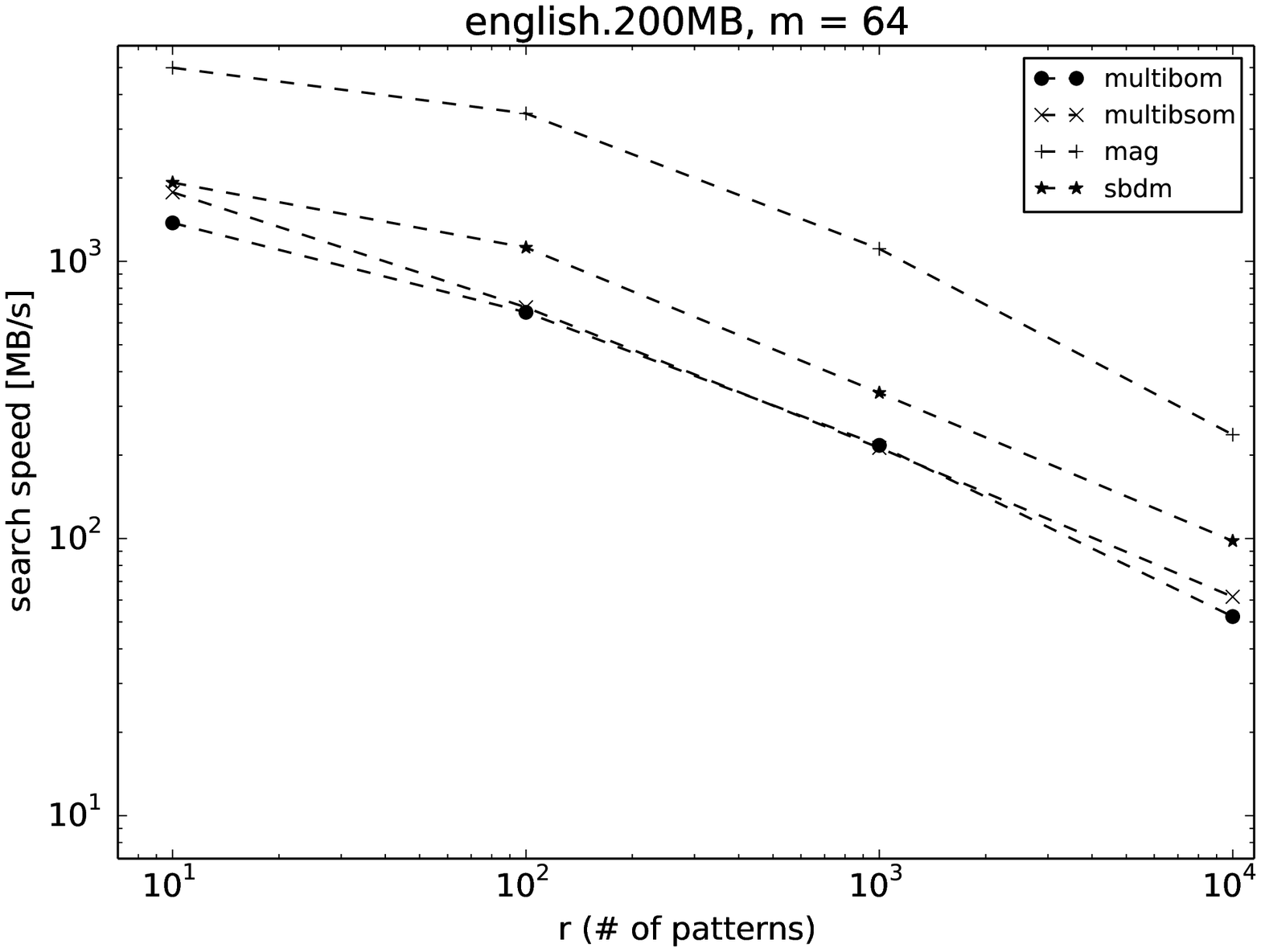}
}
\caption[Results]
{\texttt{english}, search speeds (MB/s) for varying $r$.}
\label{fig:english_varying_r}
\end{figure}

In Fig.~\ref{fig:english_varying_r} we show the results of 
all the listed algorithms on \texttt{english}, with a fixed pattern length 
$m$ and growing number of patterns $r$.
The used pattern lengths (one for each plot) are $\{8, 16, 32, 64\}$.
Note that some algorithms (or rather their available 
implementations) cannot handle longer patterns ($m = 64$).
MAG dominates for longer patterns (32, 64) and its performance is mixed 
for $m = 8$ and $m = 16$.
As expected, for all algorithms the search speed deteriorates with the 
number of patterns, and for $r = 10,000$ and relatively long patterns ($m = 32$) 
only MAG slightly exceeds 100\,MB/s (the worst ones here, SOG and RK, are 
10 times slower).

\begin{figure}[pt]
\centerline{
\includegraphics[width=0.49\textwidth,scale=1.0]{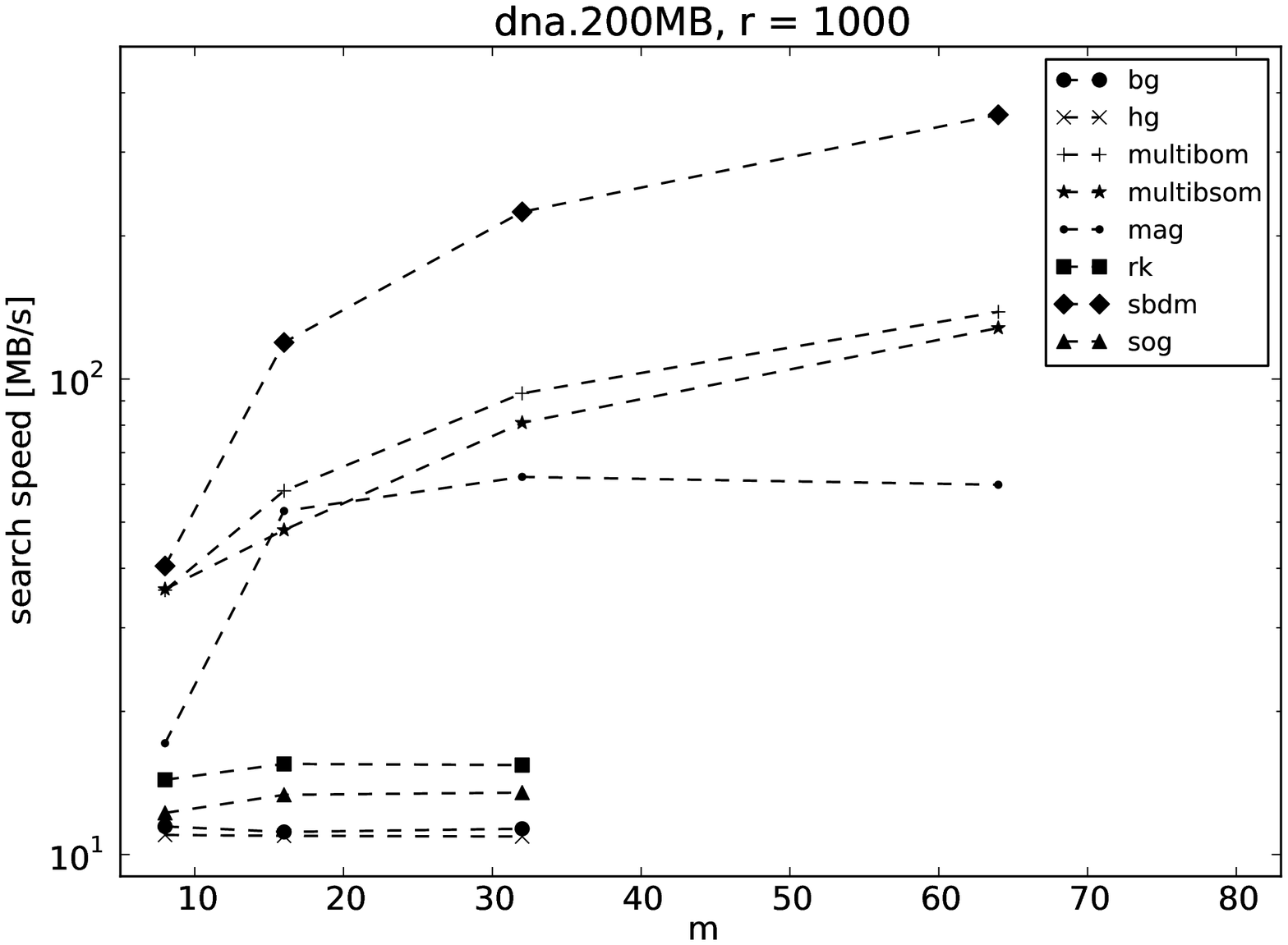}
\includegraphics[width=0.49\textwidth,scale=1.0]{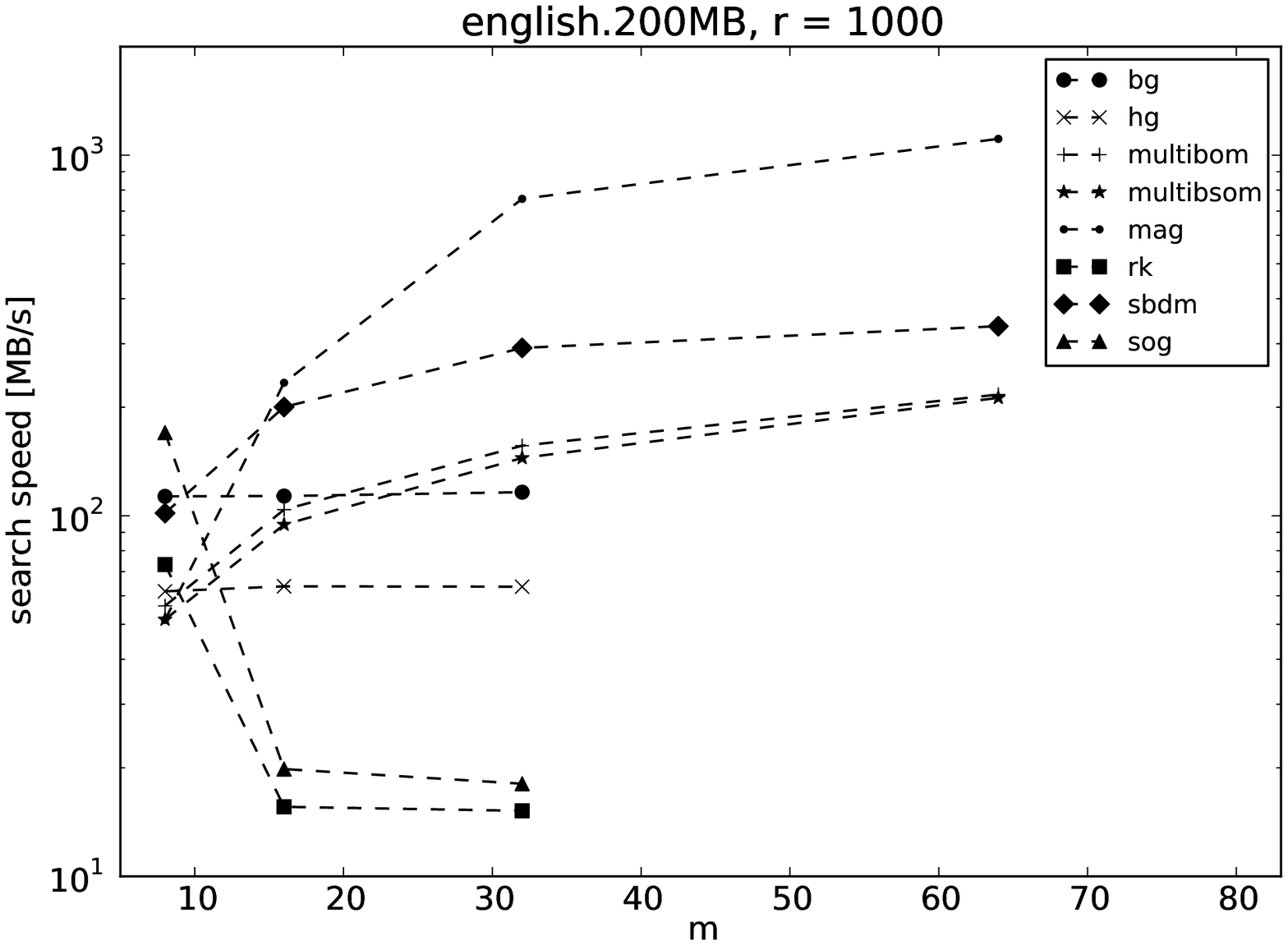}
}
\centerline{
\includegraphics[width=0.49\textwidth,scale=1.0]{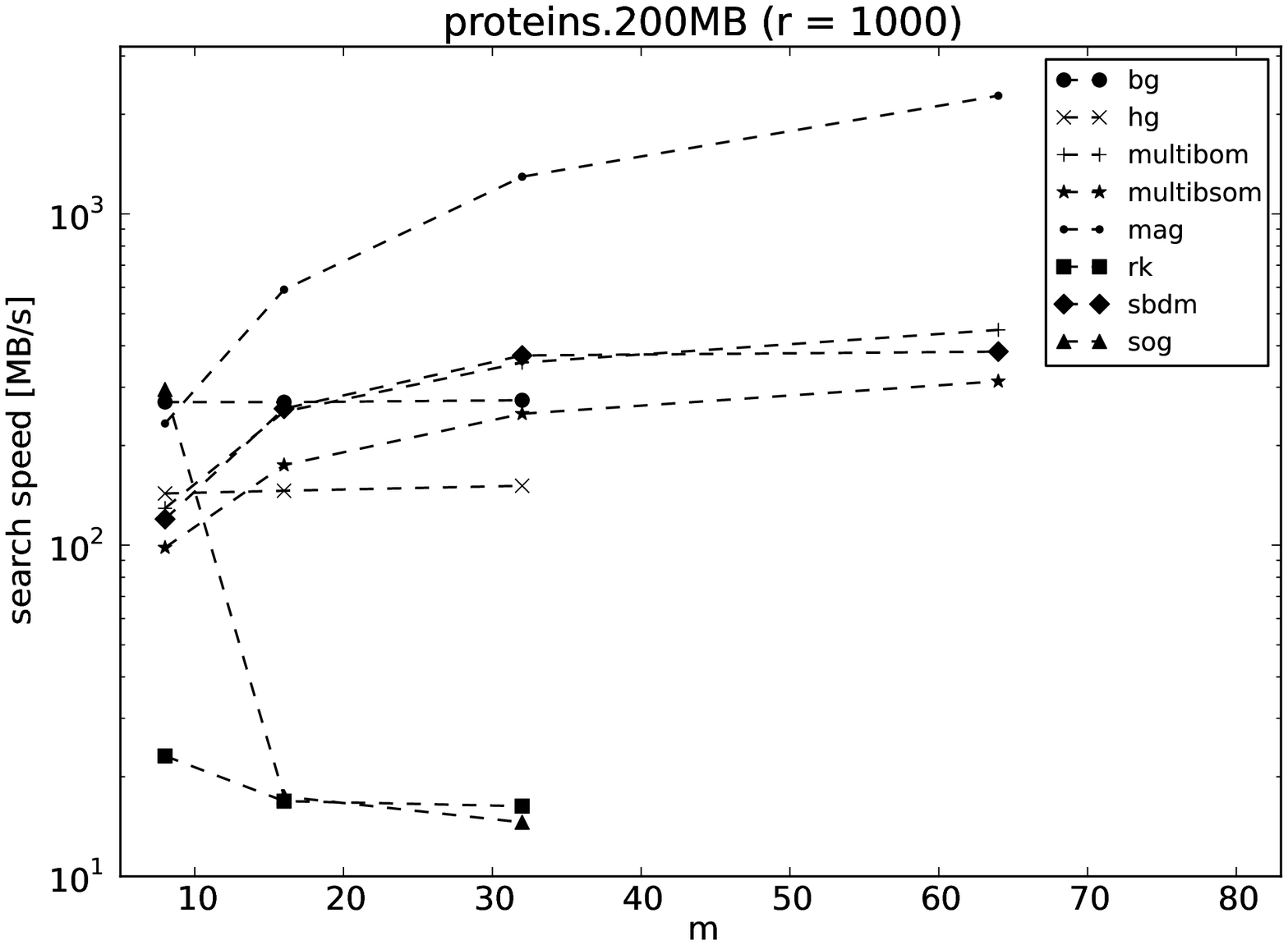}
}
\caption[Results]
{Search speeds for $r = 1000$ and varying $m$.}
\label{fig:3datasets_varying_m}
\end{figure}

In Fig.~\ref{fig:3datasets_varying_m} the number of patterns is fixed (1000), 
but $m$ grows.
MAG usually wins on \texttt{english} and \texttt{proteins} (except for 
the shortest patterns), yet is dominated by a few algorithms on \texttt{dna}.
Overall, in the experiments the toughest competitor to MAG was SBDM, 
but in some cases the winner was SOG.

\section{Conclusions and future work}

Multiple string matching is one of the most exploited problems in stringology. 
It is hard to find really novel ideas for this idea, and our work can also 
be seen as a new and quite successful combination of known building bricks.
The presented algorithm, MAG, usually wins with its competitors on the 
datasets \texttt{english} and \texttt{proteins}, but not on \texttt{dna}.
The problem is imperfect alphabet mapping, which will be addressed in future 
research.

Apart from the mentioned issue, there are a number of interesting questions 
that we can pose here.
We analytically showed that the presented approach is sublinear on average, 
yet not average optimal. 
Therefore, is it possible to choose the algorithm's parameters in order to 
reach average optimality (for $m = O(w)$)?

Real computers nowadays have a hierarchy of caches in their CPU-related 
architecture and it could be interesting to apply the I/O model 
(or cache-obvious model) for the multiple pattern matching problem.
The cache efficiency issue may be crucial for very large pattern sets.

The underexplored power of the SIMD instructions also seems to offer 
great opportunities, especially for bit-parallel algorithms.

It was reported that dense codes (e.g., ETDC) for words or $q$-grams 
not only serve for compressing data (texts), but also enable faster 
pattern searches.
Multiple pattern searching over such compressed data seems unexplored
yet and it is interesting to apply our algorithm for this scenario
(our preliminary results are rather promising).

\bibliographystyle{psc}          
\bibliography{multi}     

\end{document}